\def\year{2022}\relax
\documentclass[letterpaper]{article} 
\usepackage{aaai22}  
\usepackage{times}  
\usepackage{helvet}  
\usepackage{courier}  
\usepackage[hyphens]{url}  
\usepackage{graphicx} 
\usepackage{natbib}  
\usepackage{caption} 
\DeclareCaptionStyle{ruled}{labelfont=normalfont,labelsep=colon,strut=off} 
\usepackage{algorithm}
\usepackage{algorithmic}
\usepackage{url}
\usepackage[utf8x]{inputenc}

\usepackage{newfloat}
\usepackage{listings}
\floatstyle{ruled}
\newfloat{listing}{tb}{lst}{}
\floatname{listing}{Listing}

\title{GlyphNet: Homoglyph domains dataset and detection using attention-based Convolutional Neural Networks}

\author{
    Akshat Gupta\textsuperscript{\rm 1},
    Laxman Singh Tomar\equalcontrib \textsuperscript{\rm 2},
    Ridhima Garg\equalcontrib \textsuperscript{\rm 3}
}

\affiliations{
    \textsuperscript{\rm 1} University of Stuttgart\\
    \textsuperscript{\rm 2} Robofied\\
    \textsuperscript{\rm 3} Friedrich Alexander Universität
    
    st180429@stud.uni-stuttgart.de, laxman.tomar@robofied.com, ridhima.garg@fau.de
}

\usepackage{bibentry}

\begin{document}
\maketitle

\begin{abstract}
Cyber attacks deceive machines into believing something that does not exist in the first place. However, there are some to which even humans fall prey. One such famous attack that attackers have used over the years to exploit the vulnerability of vision is known to be a Homoglyph attack. It employs a primary yet effective mechanism to create illegitimate domains that are hard to differentiate from legit ones. Moreover, as the difference is pretty indistinguishable for a user to notice, they cannot stop themselves from clicking on these homoglyph domain names.
In many cases, that results in either information theft or malware attack on their systems. Existing approaches use simple, string-based comparison techniques applied in primary language-based tasks. Although they are impactful to some extent, they usually fail because they are not robust to different types of homoglyphs and are computationally not feasible because of their time requirement proportional to the string's length.
Similarly, neural network-based approaches are employed to determine real domain strings from fake ones. Nevertheless, the problem with both methods is that they require paired sequences of real and fake domain strings to work with, which is often not the case in the real world, as the attacker only sends the illegitimate or homoglyph domain to the vulnerable user. Therefore, existing approaches are not suitable for practical scenarios in the real world. In our work, we created GlyphNet, an image dataset that contains 4M domains, both real and homoglyphs. Additionally, we introduce a baseline method for a homoglyph attack detection system using an attention-based convolutional Neural Network. We show that our model can reach state-of-the-art accuracy in detecting homoglyph attacks with a 0.93 AUC on our dataset. 
\end{abstract}


\section{Introduction}
In cyber security, attackers employ different attacks to infiltrate our systems and networks, with the objective varying from stealing crucial information to inflicting system damage. One such deceptive attack is the homoglyph attack ~\cite{woodbridge2018detecting}, which involves an attacker trying to fool humans and computer systems by using characters and symbols that may appear visually similar to characters used in real domain and process names but are different. For example, a typical homoglyph attack may involve changing ``d" to ``cl", ``o" to ``$\theta$", and ``l" to ``1". 

\begin{figure}[ht]
    \includegraphics[width=\columnwidth]{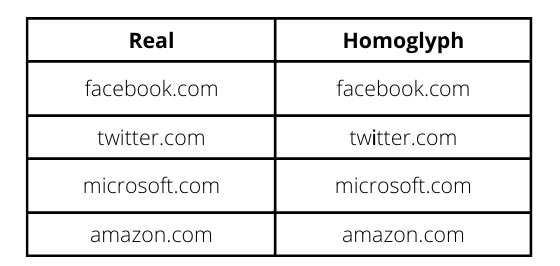}
    \caption{example of a real domain and their homoglyphs}
    \label{fig:fig1}
\end{figure}

Some of the above substitutions can be difficult for the naked eye to detect, as shown in  Figure\ref{fig:fig1}, It would mean that users would be easily susceptible to clicking on the homoglyph links, more so when navigating from one website to another. The problems arising from such an attack are of two types: a) Deceiving humans to believe that an illegitimate domain name is real by fooling the users, resulting in users using fake webpages as if they were the real ones. b) Create fake academic documents and papers by changing the real strings with homoglyphs to deceive plagiarism detection tools such as Grammarly.com \\

\begin{figure}
    \includegraphics[width=\columnwidth]{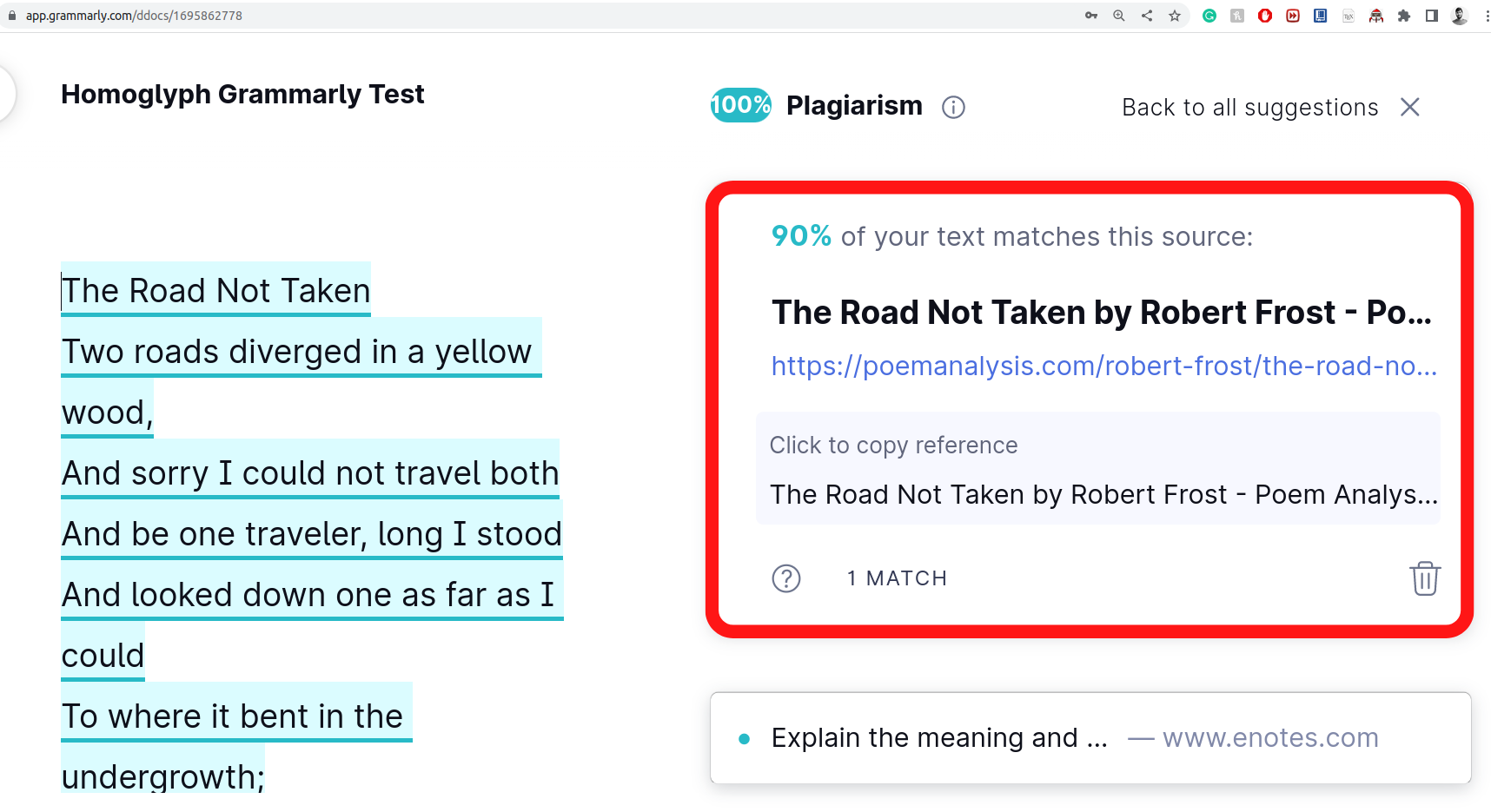}
    \caption{Real text on Grammarly plagiarism detector}
    \label{fig2}
\end{figure}

\begin{figure}
    \includegraphics[width=\columnwidth]{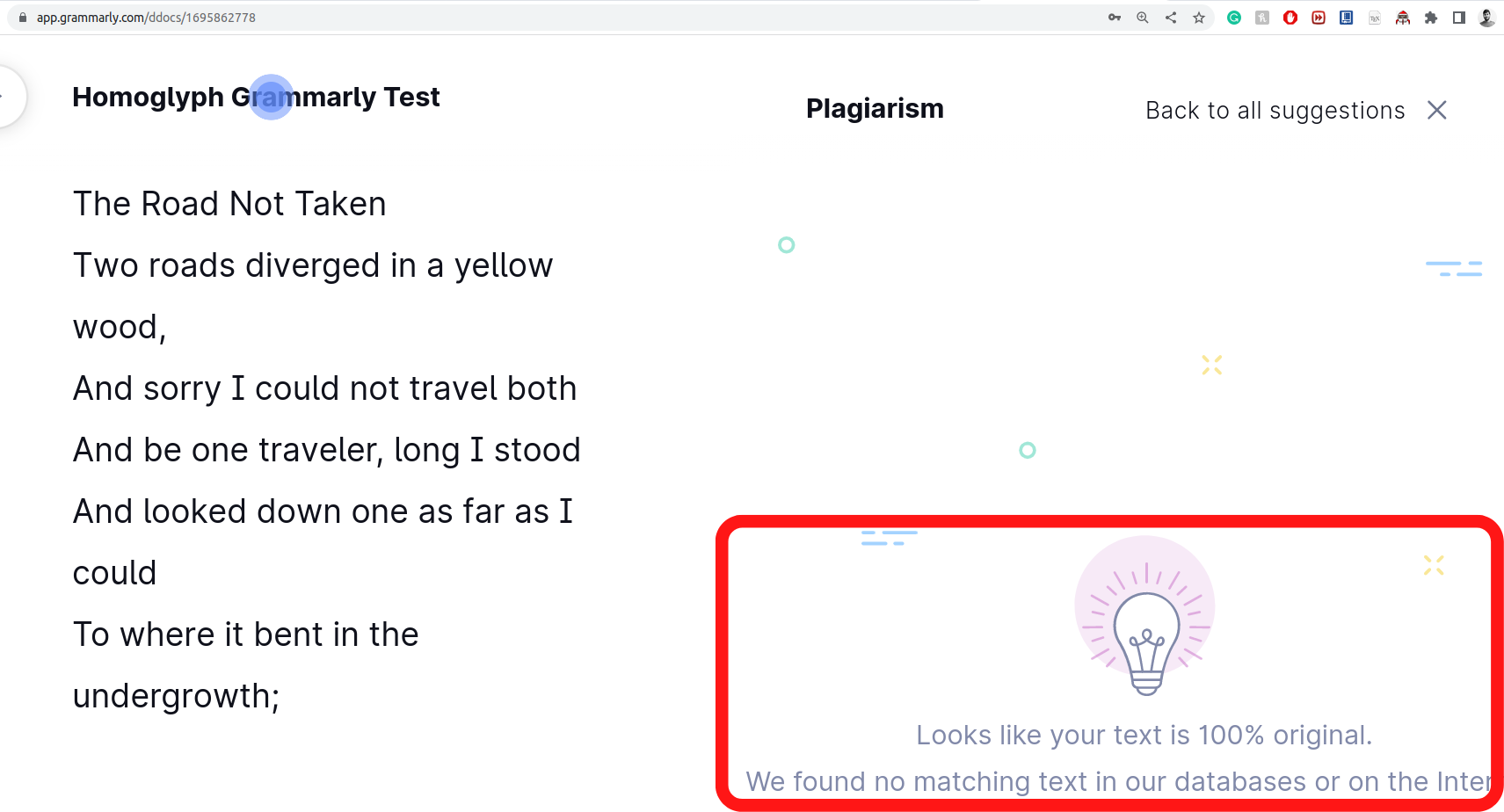}
    \caption{Homoglyph text on Grammarly plagiarism detector}
    \label{fig3}
\end{figure}

Both types of problems are hard to detect and hence require robust methods to identify an attack before it causes any information breach. Previous approaches mainly used methods from comparative algorithms such as edit distance to identify homoglyph attacks from legit strings\cite{damerau1964technique}. Any domain name string that returned an edit distance beyond an acceptable threshold was considered a homoglyph. Edit distance covers simple approaches like insertion, deletion, transposition, swapping, and substitution. Due to this shortcoming, a slight change to the illegitimate domain name can easily bypass it as quickly as a real one. Now, a slightly better version of it was proposed, which was called Visual Edit Distance\cite{ristad1998learning}. It proposes to have a particular edit distance for the visual similarity of the two domain name strings. However, these methods were more relevant in academia and had negligible prevalence in the real world. A homoglyph attack differs from a phishing attack because domain names in the former are hardly distinguishable but can be apparent in the latter.

We have taken the famous poem "The Road Not Taken" by Robert Frost to demonstrate this concept. In Figure 2, we have taken the poem text and run Grammarly's Plagiarism Detector tool. It reports 100\% plagiarism, which is correct but later when we passed the homoglyphed version of the same text, it reports the text to be 100\% unique, as shown in Figure 3. This proves that even today's state-of-the-art systems cannot effectively deal with texts comprising homoglyphs.

Recently, Microsoft obtained a court order to remove many fraudulent "homoglyph" domains used to conduct fraud and pose as Office 365 users.\cite{Microsof97:online} following a customer complaint about a business email compromise attack, Microsoft conducted an investigation and discovered that the unidentified criminal organization responsible for the attack also created 17 other malicious domains, which were combined with the customer credentials that had been stolen to gain unauthorized access to Office 365 accounts and monitor the contacts of the customers.

Microsoft stated that the cybercriminals have caused and continue to cause irreparable injury to Microsoft, its customers, and the general public. The complaint also stated that the cybercriminals have illegally accessed customer accounts, monitored customer email traffic, gathered information on pending financial transactions, and criminally impersonated [Office 365] customers.

According to studies, this attack hit $71\%$ organizations in $2021$. Sixty-two countries people were the subject of a massive cyberattack last year.

In this research, we aim to create a data set that can help expand research on homogylph attacks. We propose to apply an attention-based Convolutional Neural Network (CNN) to detect homoglyphs without the need to provide paired data. Additionally, our model achieves a significant performance boost compared to other approaches due to its architectural design. Our method can be applied directly as an API or web service for an organization to check a domain or process name before accessing it. For evaluation, we compared our performance with other baselines and found that our model outperforms them. Moreover, our approach also addresses the problem of unpaired data setting, which is often the case in the real world. 

The major contributions of our research are as follows:
\begin{enumerate}

\item Created a benchmark dataset of 4 million real and homoglyph domain images based on known homoglyph patterns and URLs. It is generated via strings from single-and dual-character noise sampled using a Gaussian distribution over a homoglyph character pool.

\item A method that uses an image dataset for detecting homoglyphs involving an attention-based convolutional neural network trained in a supervised fashion achieves a better AUC score than the other existing baselines.
\end{enumerate} 

The paper's organization starts by introducing the problem faced by existing approaches to detect homoglyph-based phishing attacks in both academia and the real world. In Related Work, we discuss the existing approaches which propound the idea of solving this problem with either string matching or Deep Learning based methods like Siamese Neural Networks and GANs. We have explained their major pitfalls in terms of generalizing capabilities and feasibility. In the Dataset section, a comprehensive description is provided of the generation of the proposed images dataset. It follows a brief description of our attention-based CNN baseline implementation. The Experimentation section describes dataset splitting, metrics used, and other settings. Later, in the Results section, we examine the results and scores obtained after the experiments conducted in the last section. Both data and baseline implementation results are validated and explained with the help of an elegant table within the same section. The following section, Discussion, presents experiments we tried that did not work. Finally, the Conclusion Section summarizes the observations and contributions.

\section{Related Work}

The work by \cite{ginsberg2018rapid} used a Siamese Neural Network to detect homoglyphs using a paired dataset. This dataset included pairs of strings; one was a real domain name, and the other was a homoglyph domain name. In their work, they converted this pair of strings into binary images that were later fed to a Siamese Neural Network\cite{koch2015siamese}. The Siamese neural network uses two identical convolutional neural networks\cite{lecun1995convolutional} to learn the visual difference between a pair of images. They were applied to domains such as healthcare, finance, and others but have recently gained popularity in cyber security.

Though\cite{lecun1995convolutional} their work showed significant improvement from previous baselines but suffered from two major pitfalls: 
\begin{enumerate}
\item In online security systems, it is impossible to provide paired data, without which these systems will not work. 

\item It cannot be used in academia due to the inability to find a paired word for each word present in a scientific article. 
\end{enumerate}

Therefore, although this approach performs well, it cannot be employed in real-world systems. 

The traditional solutions to prevent homoglyph attacks were inspired by genomics\cite{lu2019homoglyph}, which proposed the idea that homoglyph domains are in string formats and, therefore, should be compared with legitimate ones to detect whether they are real or not. Edit Distance\cite{ristad1998learning} is the measure of the minimum number of operations required to transform one sequence (domain or process name string in our case) into another. If the value exceeds an acceptable threshold, it should predict as homoglyph. This looks effective but not when giving it a second thought. The reason is that in cases like "google.com" and "go0gle.com", edit distance would return only '1' which does not look so threatening but is a homoglyph domain name. Furthermore, a paired sequence of strings is required to make comparisons, which would not be the case if it was a homoglyph of a new domain name. Finally, in the real world, this approach lacked severely good results.

Phishing attacks\cite{hong2012state} should not be confused with homoglyph attacks. Phishing is an attack involving the attacker sending homoglyph/false/fake links that appear to be coming from a trusted vendor. It leads to information compromise\cite{helms2000risk}, data breaches\cite{cheng2017enterprise}, and financial fraud\cite{reurink2018financial}. The difference between Phishing and Homoglyphs is that the former uses tricks such as longer, bad, and misspelled domain names and URLs\cite{ma2009identifying} to fool people. At the same time, the latter takes advantage of the inability to differ in visually similar but different domain name strings. Thus, it is required to create better solutions for homoglyph detection.

\subsection{Siamese Neural Networks}
\begin{figure}
    \includegraphics[width=\columnwidth]{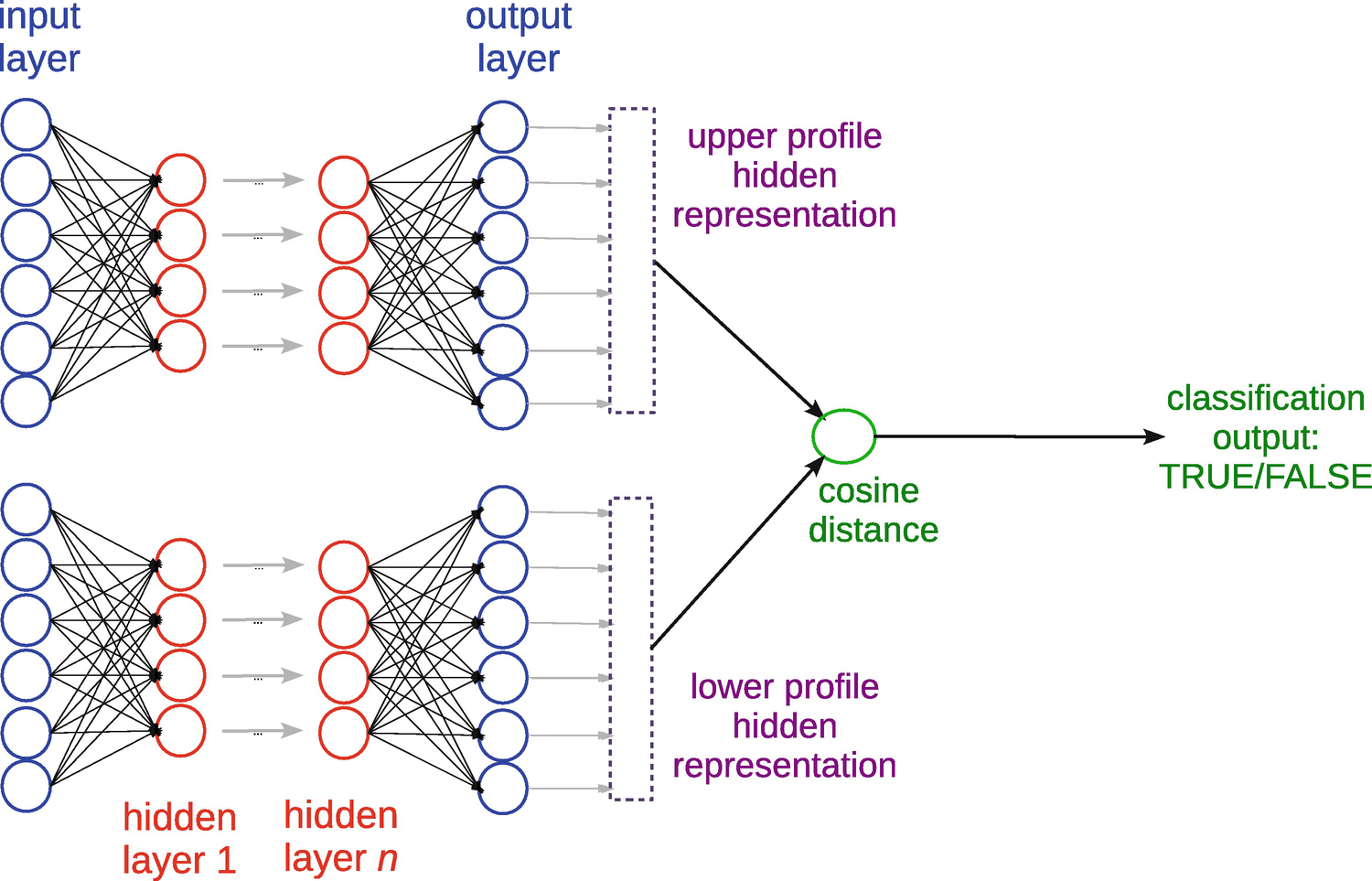}
	\caption{Siamese neural network architecture\cite{woodbridge2018detecting}}
	\label{fig0}
\end{figure}

The Siamese neural network architecture is proposed to detect homoglyphs using a paired data set. This dataset included pairs of strings, one was a real domain name, and the other was a homoglyph domain name. Each instance was a tuple that contained a real domain string, a homoglyph domain string, and a score that denotes whether the second element is a valid homoglyph of the first or not as part of its elements. In their work, they converted this pair of strings into binary images, images that were later fed to a Siamese Neural Network. However, we observed a significant difference while reproducing the results in our dataset.

\subsection{PhishGANs}
\begin{figure}
    \includegraphics[width=\columnwidth]{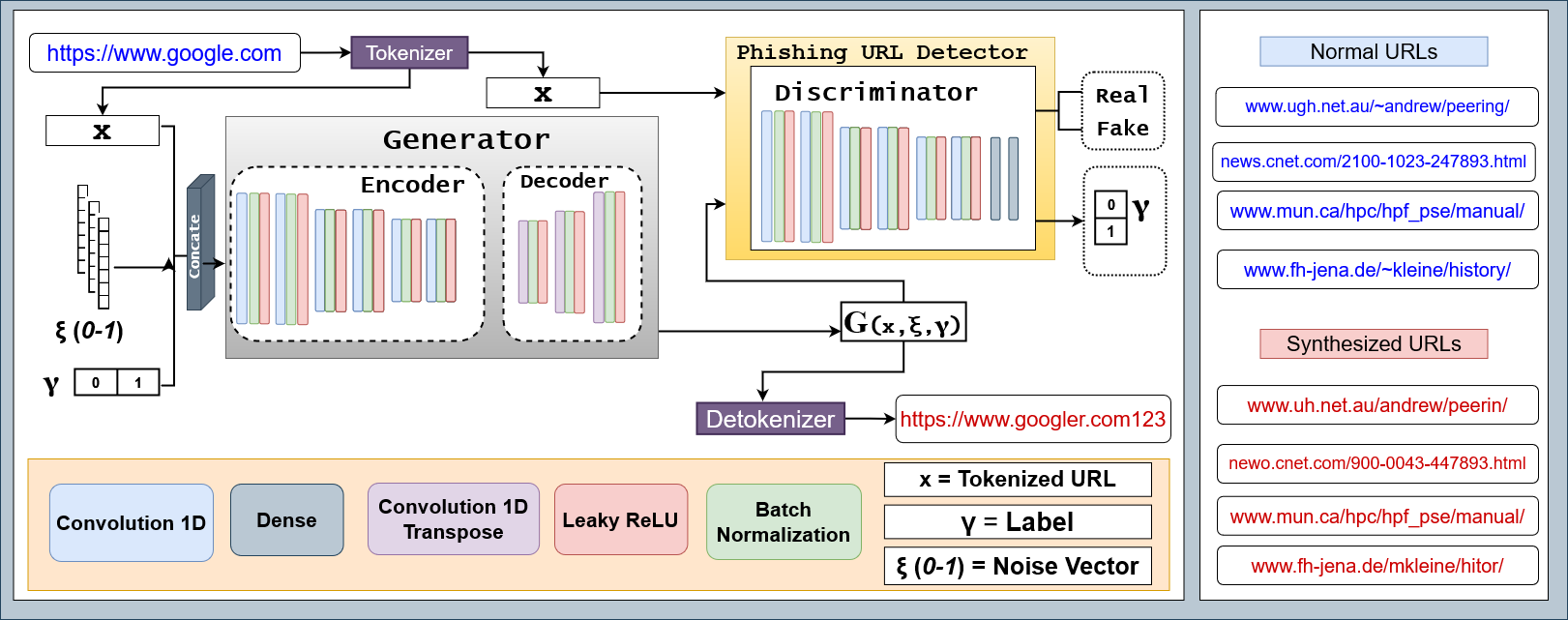}
	\caption{PhishGAN architecture\cite{sern2020phishgan}}
	\label{fig4}
\end{figure}

Approaches such as Siamese Neural Networks suffered severely in terms of performance due to lack of data, as they only had close to $91k$ real domain images. As a remedial solution, we were required to produce comprehensive data to train our models well. Recently, Lee Joon Sern et al. proposed PhishGANs to generate synthetic data\cite{sern2020phishgan}. They discussed creating a generative adversarial network\cite{goodfellow2014generative} that aimed to create images similar to real domain names to increase existing data sets. PhishGANs\cite{sern2020phishgan} being a GAN\cite{goodfellow2014generative} involved a generator and a discriminator, both trained in an adversarial fashion such that the generator is trained well enough to produce images similar to those of a real domain which the discriminator cannot detect. Later, these images were fed to a different network for binary classification aimed at distinguishing real domain names from homoglyphs. They used UNet\cite{ronneberger2015u} architecture as a generator using a custom loss function called a dot product loss. The PhishGANs\cite{sern2020phishgan} were trained similarly to how Pix2Pix\cite{isola2017image} is trained. Later, for classification purposes, a new architecture was defined and called homoglyph identifier (HI) using CNN\cite{lecun1995convolutional} as an encoder using a triplet loss function \cite{hoffer2015deep} as input, the positive domain (google.com), the anchor domain (go0gle.com) and the negative domain (apple.com). On some popular domains, such as youtube.com and facebook.com, HI achieved an accuracy of roughly $0.81$ while testing their homoglyphs. On an unseen domain, HI achieved an accuracy of $0.76$ while feeding it back again in PhishGANs\cite{sern2020phishgan} and generating its homoglyphs, and later training on them, which helped detect its homoglyphs using $0.86$ accuracy. Although the idea of generating synthetic data using GANs\cite{goodfellow2014generative} looks promising and intriguing but is not motivating when it comes to real-world usage. GANs\cite{goodfellow2014generative}, in general, is one of the trickiest architectures in Deep Learning\cite{lecun2015deep} since their advent and are often found to have issues while training in the real world, which is not the case in the constrained environment of academia. It is common to encounter issues like problems in convergence, generator oscillating between generating specific examples in the domain, and multiple inputs resulting in generating the same output. Also, the performance increase was not drastic enough to compel us for its usage.

\section{Dataset}
The work by \cite{woodbridge2018detecting} proposed their custom paired data set that comprises $91k$ real domains and $900k$ homoglyphs. Each real domain is used to generate its respective homoglyphs. Each point in this dataset is a three-element tuple denoting domain, homoglyph, and score. Here, if the value of the score is $1.0$, then it is a valid homoglyph of the real domain. The real-world data limitation to Homoglpyh-based attacks is the lack of publicly available data sets. \\

\textbf{Proposed dataset: GlyphNet} \\
We have proposed a dataset consisting of real and homoglyph domains. To generate homoglyph domains, real domains are needed. We have obtained domains from the Domains Project\cite{tb0hdand18:online}. This repository is one of the largest collections of publicly available active domains. The entire repository comprises 500M domains, and we restricted our work to 2M domains due to hardware restrictions. \\

\textbf{Homoglyph Creation Algorithm} \\
Homoglyph Generation is an important task, as one needs to ensure enough randomness to make it appear real and keep the process simple enough to fool the target. Publicly available tools like dnstwist\cite{elceefdn36:online} replace every character in the real input domain with their respective glyphs. It generates poor homoglyphs for the large part because it relies on paired data which is not fit to serve the purpose practically. We created our novel algorithm for the generation of homoglyph domains to ensure that real homoglyphs are generated with randomness and closeness. To achieve this, we sample homoglyph noise characters using Gaussian sampling\cite{boor1999gaussian} from the glyph pool. We used 1M real domains to generate $2M$ homoglyphs with a single glyph character and introduce diversity in our dataset; we reran this algorithm on the remaining 1M real domains to generate homoglyph domains with two character glyphs. Finally, we have the 4M real and homoglyph domains. \\

\textbf{Image Generation} \\
Homoglyph attacks exploit the weakness of human vision to differentiate real from homoglyph domain names. From a visual perspective, we are interested in learning the visual characteristics of real and homoglyph domain names. To do so, we rendered images from the real and homoglyph strings generated via our algorithm. We have used ARIAL Typeface as our chosen font, a $28$ font size, on a black background with white text from the middle left of the image; the image size is $150\times150$. 

\begin{table*}[!ht]
    \centering
    \begin{tabular}{|c|c|c|c|} \hline
         \textbf{Dataset Name} & \textbf{Real} & \textbf{Homoglyph} & \textbf{Total} \\ \hline
         Domain and Process Strings\cite{woodbridge2018detecting}& $90k$ & $900k$ & $990k$ \\ \hline
         Similar and Dissimilar Pairs\cite{majumder2020convolutional} & $2257$ & $2257$ & $4514$ \\ \hline
         \textbf{GlyphNET (Ours)} & \textbf{$2000k$} & \textbf{$2000k$} & \textbf{$4000k$}\\ \hline
    \end{tabular}
    \caption{Dataset comparison}
    \label{tab:my_label}
\end{table*}

\begin{figure}
    \includegraphics[width=\columnwidth]{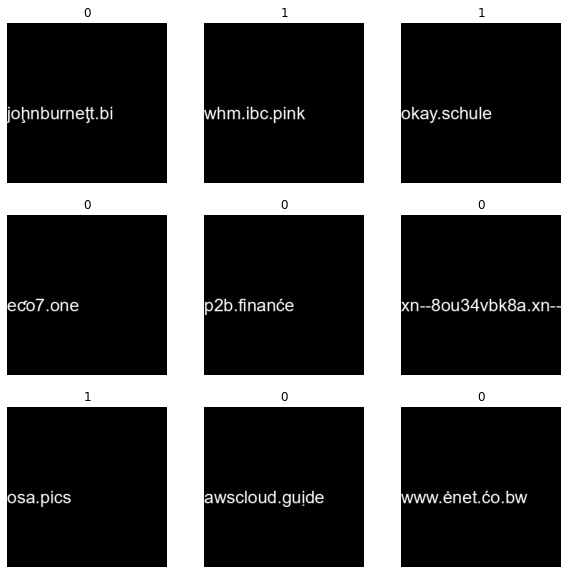}
    \caption{Rendered images from the dataset, $0$; homoglyph domain and, $1$; real domain}
    \label{fig5}
\end{figure}

\section{Methodology}

This section presents our approach to building an end-to-end homoglyph detection system. We build on attention-based\cite{bahdanau2014neural}\cite{vaswani2017attention} convolutional neural network\cite{lecun1995convolutional} that aims to exploit the visual dissimilarity between real and homoglyph domain names. The architecture of our model is shown in Figure 7 and Figure 8.

The rendered images are then used as input to the CNN to learn the desired visual feature information. The model consists of four Conv2D layers to learn visual information such as edges, curves, and strokes. Each convolutional layer is paired with a max-pooling layer to perform dimensionality reduction on the learned features. This model is developed in keras\cite{chollet2015keras}. Each convolution block is followed by a convolutional block attention module (CBAM), as described in the following.

\begin{figure*}[!ht]
  \includegraphics[width=\textwidth,height=6cm]{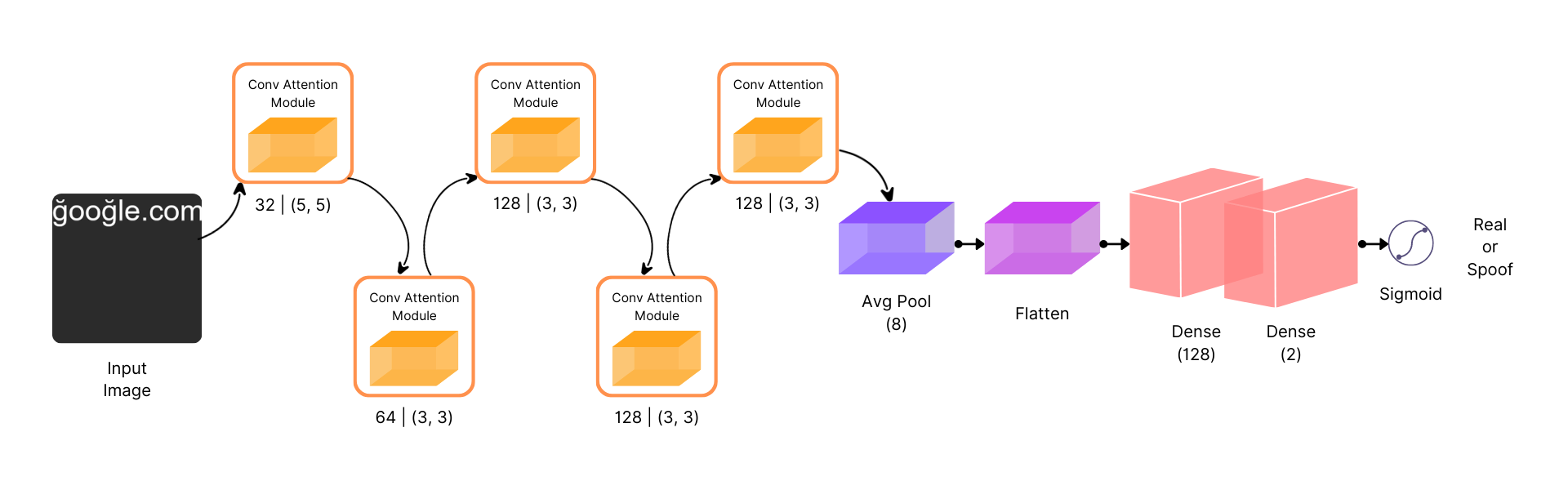}
  \caption{Our neural network architecture}
\end{figure*}

\begin{figure*}[!ht]
  \includegraphics[width=\textwidth,height=6cm]{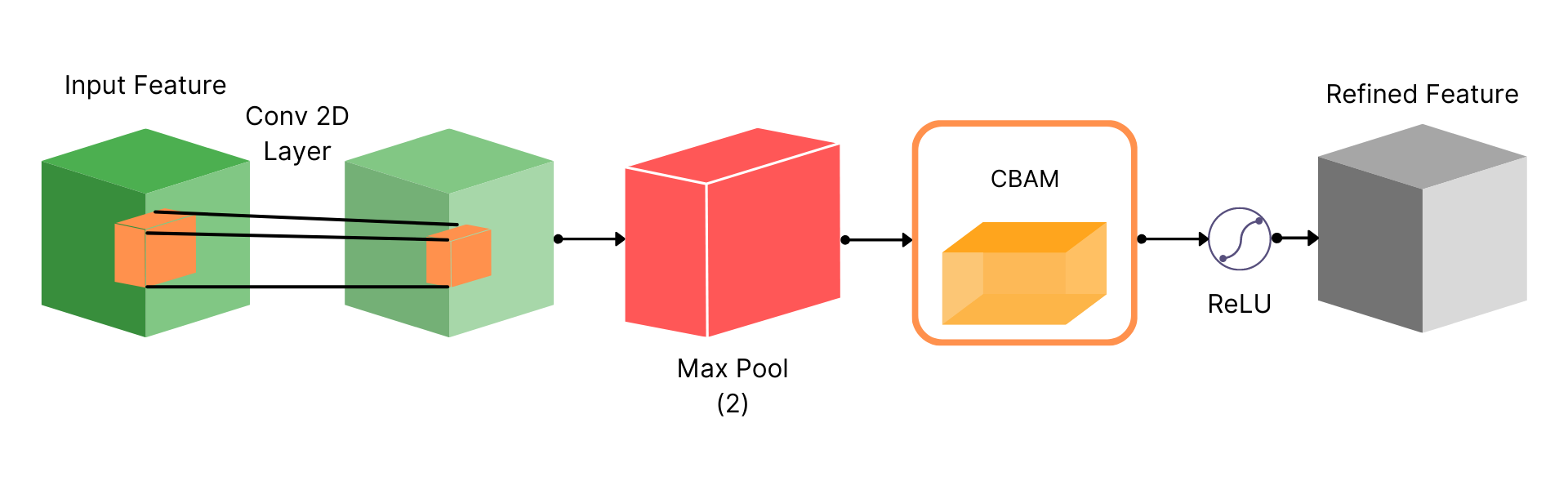}
  \caption{Zoom in view of conv-attention module}
\end{figure*}

Attention processes boost the strength of representation by focusing on essential traits and suppressing unneeded ones. It uses the feed-forward convolutional neural network's CBAM, a specific and efficient attention module. Given a preliminary feature map, the module successively infers attention maps along the channel and spatial dimensions. It then multiplies the attention maps by the preliminary feature map to achieve adaptive feature refinement. The overall attention process is summarized as follows:

\begin{center}
$F'=M_c(F) \otimes F,$ \\
$F''=M_s(F') \otimes F'$,
\end{center}
\begin{enumerate}
    \item Given an intermediate feature map $ F \in \mathcal{R}^{C \times H \times W}$ as input. $C$ represents a number of channels, $H$ and $W$ represent the height and width of the feature map $F$ respectively. 
    \item CBAM sequentially infers a 1D channel attention map $M_c \in \mathcal{R}^{C \times 1 \times 1 }$ 
    \item And a 2D spatial attention map $M_s \in \mathcal{R}^{1 \times H \times W}$ 
    \item $\otimes$ element-wise multiplication
\end{enumerate}

For the sake of non-linearity, the RELU activation function is used. 

\begin{table*}[!ht]
    \centering
    \begin{tabular}{|c|c|c|c|c|c|} \hline
         \textbf{Architecture}& \textbf{Accuracy} & \textbf{Precision} & \textbf{Recall} & \textbf{F1-score} & \textbf{AUC}\\ \hline
         Siamese CNN\cite{woodbridge2018detecting} & $0.79$ & $0.78$ & $0.71$ & $0.74$ & $0.78$\\ \hline
         Ensemble CNN\cite{majumder2020convolutional} & $0.83$ & $0.82$ & $0.79$ & $0.80$ & $0.83$\\ \hline
         PhishGAN\cite{sern2020phishgan} & $0.71$ & $0.74$ & $0.65$ & $0.69$ & $0.71$\\ \hline
         \textbf{Attention CNN (Ours)} & \textbf{$0.93$} & \textbf{$0.93$} & \textbf{$0.93$} & \textbf{$0.93$} & \textbf{$0.93$}\\ \hline
    \end{tabular}
    \caption{Model performance comparison on our dataset}
    \label{tab:results}
\end{table*} 

\section{Experimentation}

\subsection{Dataset and Metrics}
We have split our dataset into three parts, train, validation, and test, with a ratio of $70:20:10$, respectively which amounts to $2.8M$, $0.8M$, and $0.4M$ images in train, validation, and test sets. Each image size is $150\times150$. 

We use accuracy for measuring the performance of the classification task. Since accuracy can sometimes be misleading in a binary classification task, especially for unbalanced data sets, we consider precision, recall, and F1 score as our evaluation metrics, even though our dataset is balanced. We have also used the AUC score to compare our solution with some other works.

\subsection{Experimental Settings}
For the training part, we used binary cross-entropy as a Loss Function. We have used RMSProp optimizer to optimize the loss obtained from the binary cross-entropy loss function, with a learning rate of $10e^{-4}$, and the network is trained for $30$ epochs with early stopping. We trained with a batch size of $256$. We evaluated the performance of our model in terms of accuracy vs. epochs and loss vs. epochs plots.

\section{Results}
We evaluated our model on two unpaired data sets for domain names. We took an input string from the dataset we created in the previous section, converted it into an image, and fed it to the model to generate a binary label. The results for the domain names are tabulated in Table 2. Out of the $400k$ test images, our model correctly categorized $372k$ images resulting in $0.93$ accuracy. Our model achieved an f1-score of $0.93$, $13$ points higher than the previous models. Our model outperforms other baselines and comparable works on the other metrics, including accuracy, precision, recall, and AUC. The performance of other models on our dataset was also below par compared with the proposed datasets in their works, signifying our dataset's variations, difficulty, and importance. 

Our dataset, code, and models are publicly available under MIT LICENSE and can be accessed from our project's GitHub repository\footnote{\url{https://github.com/Akshat4112/Glyphnet}}

\section{Discussion}
We now discuss some interesting observations and experiments which did not work and possible explanations regarding them.

\subsection{Using only Grayscale Images}
During the image rendering phase, where we generated images from the data set containing real and homoglyph domains, we experimented with generating colored images instead of grayscale ones. We used ($73$, $109$, $137$) as the background color while ($255$,$255$,$0$) as the color of the text to be written. However, the network trained from these colored images was always found to be underperforming the network trained on grayscale images. One possible reason might be that the grayscale involves black and white as two colors, which are two extremes. Hence, it preserves the difference in adjoining pixels at the periphery of the letter and background pixels.
Meanwhile, the colors though appearing to us as distinctly different, suffered to preserve the difference when later passed through resizing operations. We perform data augmentation on our data and later train our network using the data generated, but it leads to a downfall in accuracy. One possible reason might be that data augmentation\cite{shorten2019survey} is used in those scenarios where we expect distinctive image features to exist, but they do not exist in the actual data set. It can be understood from a Cat vs. Dogs example. Usually, data sets contain cats and dogs in limited positions in the pictures, so our model fails to recognize some of the real images. The reason is that in the real world, either a cat or a dog might turn their heads and might be sitting in different postures, which makes it difficult for our model to locate distinctive features like whiskers and pointy ears in cats and tongues in case of dogs in the absence of large amounts of data catering these considerations. Therefore, to mimic such behavior, Data Augmentation is used, which helps to create all these different types of images. However, in our case, using it leads to flipped characters, and rotated images lead to anchor and tilde signs over letters going in different directions, which is not the case with real-world strings. Therefore, data augmentation was, in fact, counterproductive for our use case.

We rendered images of sizes $256\times256$ during the image generation phase. Apart from the image size $256\times256$, at which we observed the best results, we tried experimenting with the following image sizes: $128\times128$, $150\times150$, $224\times224$, and $512\times512$. The smaller the image size, the more performance degradation there is relative to it. An increase in size did not lead to any significant improvement but increased the training time of the model. Hence, we use $256\times256$ image size.

\subsection{Building Model without Transfer Learning}
We train a base network on a base dataset and task and then reuse the learned features or transfer them to a second target network to be trained on a target dataset. This process will tend to work if the features are general, meaning suitable for both base and target tasks rather than specific to the base task. We performed experiments with transfer learning\cite{pan2009survey} by incorporating networks like VGG16\cite{simonyan2014very}, Resnet18\cite{he2016deep}, Resnet34, Resnet50, Wide ResNet-101-2, ResNeXt-50-32x4d and ResNeXt-101-32x8d which were trained on ImageNet\cite{deng2009imagenet} dataset. Our experiments did not obtain good accuracy using these architectures, either pre-trained or from scratch.

There are two possible reasons: 

1) Large number of hidden layers: These architectures have many hidden layers ranging from $16$ up to $100$. The deeper the network, the more it tries to aggregate the learned features to create high-level features. It works well in images of real-world entities, but in our context, it does not help as these are just images generated from strings. Going further deep into the network makes the network lose all the subtle features from parts of strings like tilde and apostrophes. It has learned to differentiate real from homoglyph strings. 

2) Pre-trained in a data set of different domains: Another reason is that these networks were pre-trained on the ImageNet dataset, which contains images from real-world entities, but does not have images similar to our problem. Hence, using a pre-trained network having weights learned from such images instead of our domain problem did not help. We obtained an accuracy of $63\%$ to $67\%$ using the above architecture.

\section{Conclusion}
In this work, we created a first-of-its-kind large-scale homoglyph phishing image dataset comprising 4M images of real and homoglyph domains. Later, we presented a baseline that relied on learning features from an attention-based convolutional neural network using our constructed data set to differentiate real domain names from homoglyph domain names to avoid homoglyph attacks. Our dataset and approach are robust because we can generalize on unseen homoglyphs as compared to other approaches which are data-dependent for every single inference, which leads it to outperform existing approaches. We believe this work is significant and provides an important benchmark to propel work in this area, and its applications would serve as a safeguard against phishing attacks in the real world. 

\bibliography{aaai22}

\end{document}